An Investigation of the Different Levels of Poverty and the Corresponding Variance in Student Academic Prosperity

Sebastian Del Barco, Erast Davidjuk
University of Colorado Denver
February 24, 2017

Abstract

Underprivileged students, especially in primary school, have shown to have less access to educational materials often resulting in general dissatisfaction in the school system and lower academic performance (Saatcioglu and Rury, 2012, p.23). The relationship between family socioeconomic status and student interest in academic endeavors, level of classroom engagement, and participation in extracurricular programs were analyzed. Socioeconomic status was categorized as below poverty level, at or above poverty level, 100 to 199 percent of poverty, and 200 percent of poverty or higher (United States Census Bureau). Student interest, engagement, and persistence were measured as a scalar quantity of three variables: never, sometimes, and often. The participation of students in extracurricular activities was also compared based on the same categories of socioeconomic status. After running the multivariate analysis of variance, it was found that there was a statistically significant variance of student academic prosperity and poverty level.

Introduction

The rise of technology throughout industry has created a need for workers that are more skilled than ever before which has decreased the amount of industry blue collar jobs available because of systemic unemployment. Student participation in school activities and in the classroom has become increasingly important for student success in the workplace and other postsecondary endeavors. However, even with such an increase in the demand for skilled workers in the labor force, there is a growing discontinuity between students that are engaged in the classroom and those students that are not which may be due to the differences in socioeconomic status. This deficit of education in underprivileged schools exacerbates the wage gap between socioeconomic groups and perpetuates the educational disparity that exists. The economic discrimination of unskilled workers and the the rising structural unemployment as a result of technological improvements is why it is paramount to find a cost-effective solution to encourage students to become more engaged academically so that these individuals have skills that incentivise companies to employ them.

Expansion of the modern economy is reliant on new job creation and employment. Education, in many cases, is a barrier for many STEM careers which are necessary for societal development. Surveying various ethnic groups and their academic achievements within the United States can provide new perspectives on this connection. Furthermore, lack of proper education materials and resources may influence the rates of student academic success.

The goal of this study was to determine if there is a statistically significant difference between the poverty level of a student and a student's academic involvement. A census was conducted on students in varying poverty rankings to find the means of the students that were engaged throughout different academic aspects such as the classroom learning experience, the individual student persistence, the amount of extracurriculars the student was involved in, and the personal student academic interest. This study looked at the variance between the means of the different poverty levels to make an inference on the relationship between poverty level and overall student academic fulfillment as measured by standards by the U.S. Census Bureau. This investigation demonstrated that there was a statistically significant difference between the different rankings of poverty level. This highlights the educational gap and what may be a leading factor in the workforce divide. As a result of the lower education in these regions with a lower poverty ranking, it may be more difficult for these students to become highly skilled workers. Consequently, companies aren't incentivized to hire people that aren't as skilled which increases the opportunity gap between income levels.

The hypothesis that was tested was that there was a statistically significant difference between the student fulfillment and the poverty levels. The null hypothesis was that there was not a statistically significant variance between the average means of the different poverty levels on the different levels of activities, engagement, persistence, and interest in school.

($H_0$: $\beta_0$=0  $H_1$: $\beta_0 \neq 0$)

($H_0$: $\beta_1$=0  $H_1$: $\beta_1 \neq 0$)

($H_0$: $\beta_2$=0  $H_1$: $\beta_2 \neq 0$)

($H_0$: $\beta_3$=0  $H_1$: $\beta_3 \neq 0$)

($H_0$: $\beta_4$=0  $H_1$: $\beta_4 \neq 0$)

The data used in this investigation was collected from the United States Census Bureau. The analysis that was performed was a multiple analysis of variance between poverty status groups and the dependent variables, student engagement, persistence, interest, and participation in extracurricular activities. The multiple analysis of variance test that was conducted with the varying levels of poverty and the student academic intent was found that there is a statistically significant variance between the means of the students within the different poverty levels at a confidence level of 95% or $\alpha= 0.05$.

Methods

The independent variables that were compared in this investigation were the four categories of poverty as defined by below poverty level, at or above poverty level, 100 to 199 percent of poverty, and 200 percent of poverty or higher. These variables were defined as interval variables were ordinal variables because the classification of poverty level has a significant difference between the order of the different classifications. These classifications were based on the dollar amount of salary that the parents of these individuals received and as such, this assumption was appropriate to make.

The dependent variables that were compared in this investigation were the student interest in school, the amount of extracurricular activities that the student was involved in, the student engagement in the classroom, and the individual student persistence. These variables were categorized as discrete variables because there are no meaningful boundaries between the variables or meaningful order between the different categories. This was an appropriate establishment of the variable categories because of the nature of the survey. Individual student persistence was categorized based on three different responses for each student which were never, sometimes, and often. To create a meaningful variable category to run the analysis, these variables were established as discrete variables. These categories remained consistent for all of the questions except for the extracurricular activity involvement. The extracurricular activity involvement was measured based on the proportion of students that participated in extracurricular activities relative to their socioeconomic background. Therefore it was appropriate to establish the extracurricular activity involvement as discrete variables.

Using a multivariate analysis of variance, the statistical significance of the variance was assessed between the different independent variables and the different dependent variables at a confidence level of $\alpha= 0.05$ or a confidence of 95%. There were multiple assumptions that were checked before the multivariate analysis of variance was performed. The first assumption was that the independent variables were scalar which was established when defining the variables to test. The second assumption was a test for independence of observations which was checked by the nature of the surveys established by the United States Census Bureau. The sample size was established to be 48,738,000 which was a sufficient sample size to assume normality of the sample. Moreover, to check if there were no multivariate or univariate outliers the Mahalanobis distance was calculated which allowed multivariate and univariate outliers to be removed from the study. Furthermore, to establish normality between the sample size a Shapiro-Wilk test was run on the sample size and the data passed this assumption test. Additionally, the homogeneity of variance-covariance matrices was accounted for by running a Levene's test of homogeneity of variance on the sample which the data passed. Lastly, multicollinearity between the sample was tested and there were no highly correlated predictors identified by the analysis of the variance inflation factors. Once all of these assumptions were tested for, it was possible to run a multivariate analysis of variance between the poverty categories and the student interest in school, amount of extracurricular activities that the student was involved in, the student engagement in the classroom, and the individual student persistence.

The Wilks' test was used as a determinant of variance in the S matrices.

$$\Lambda = \frac{|S_{Error}|}{|S_{Effect} + S_{Error}|}$$

The estimated F statistic can be found with the equation

$$F_{Estimate}(df_1, df_2) = (\frac{1-y}{y})(\frac{df_2}{df_1})$$

The p value that was used is based on the amount of determinant variables that existed in the test that is being run.
$p = \#DVs$

The Shapiro Wilk's test statistic was found by the following equation:

$$W = \frac{(\sum_{i=1}^{n} a_i x_i)^2}{\sum_{i=1}^{n}(x_i - \bar{x})^2}$$

Where
$x_i$ is the smallest number in the sample.
$\bar{x}$ is the sample mean.
$a_i$ is a constant that is given by $= \frac{m^T V^{-1}}{(m^T V^{-1} V^{-1} m)^{1/2}}$

This was used as a determinant of normality which can be found by the test statistic.

The Levene Test was used to assess the variances between the different variables in this sample which is a measure of homoscedasticity.

The Levene Test statistic W was found by the following equation:

$$W = \frac{(N-K)}{(k-1)} \cdot \frac{\sum_{i=1}^{k} N_i(Z_i - Z)^2}{\sum_{i=1}^{k}\sum_{j=1}^{N_i}(Z_{ij} - Z_i)^2}$$

Where:
K is the amount of groups sampled.
$N_i$ is the amount of cases observed.
N is the number of cases in all of the groups.
$N_{ij}$ is the numerical that is measured from the sample.

$$Z_i = \frac{1}{N} \sum_{j=1}^{N_i} Z_{ij} \text{ which is the mean of } Z \text{ for the sample.}$$

$$Z = \frac{1}{N} \sum_{i}^{k}\sum_{j=1}^{N_i} Z_{ij} \text{ is the mean of all } Z_{ij} \text{ in the sample.}$$

It is important to find the error of the determinants which is found by the following:

$$df_{Error} = n_{L1} \cdot n_{L2} \cdot n_{L3} \cdot n_{L4}(n_{DV} - 1)$$

Next the effect of the determinants is calculated based on the former equation:

$$df_{Effect} = (IV_1 - 1)(IV_2 - 1)(IV_3 - 1)(IV_4 - 1)$$

The first determinant of was calculated based on the following equation:

$$df_1 = p(df_{Effect})$$

The second determinant is based on the following relationship between the p value established and the error of both determinants.

$$df_2 = s\left[(df_{Error}) - \frac{p - df_{Effect} + 1}{2}\right] - \left[\frac{p(df_{Effect}) - 2}{2}\right]$$

The standard deviation of the probability distribution that is used in this multiple analysis of variance is modeled by this equation:

$$s = \sqrt{\frac{p^2(df_{Effect})^2 - 4}{p^2 + (df_{Effect})^2 - 5}}$$

The association strength between the variables was measured with the most favorable combination of linear DVs.

$$\eta^2 = 1 - \Lambda$$

Once the initial assumptions were tested, the data set passed these assumptions ,and outliers were removed from the data set it was possible to the the multi variance analysis of variance on the data set.

Sum of Squares$_{total}$= Sum of Squares$_1$+ Sum of Squares$_2$+ Sum of Squares$_3$+ Sum of Squares$_4$

This can be expressed by the following:

$$\sum_i \sum_j (Y_{ij} - GM)^2 = n \sum_j (\overline{Y_j} - GM)^2 + \sum_i \sum_j (Y_{ij} - \overline{Y_{ij}})^2$$

However, because there are multiple independent variables in this investigation the standard equation for multiple independent variables with a multivariate analysis needs to be applied. There are four independent variables that are compared in this study which yields the following equation.

$$n_{kmlx} \sum_k \sum_m \sum_l \sum_x (IV_1 IV_2 IV_3 IV_4 - GM)^2 = n_k \sum_k (IV_{1k} - GM)^2 + n_m \sum_m (IV_{2m} - GM)^2 + n_l \sum_l (IV_{3l} - GM)^2 +$$

$$n_x \sum_x (IV_{4x} - GM)^2 + \left[ n_{km} \sum_k \sum_m \sum_l \sum_x (IV_1 IV_2 IV_3 IV_4 - GM)^2 \right] [- n_k \sum_k (IV_{1k} - GM)^2 - n_m \sum_m (IV_{2m} - GM)^2$$

$$- n_l \sum_l (IV_{3l} - GM)^2 - n_x \sum_x (IV_{4x} - GM)^2]$$

This is the equation for the MANOVA that was run with the four categories of poverty as defined by below poverty level, at or above poverty level, 100 to 199 percent of poverty, and 200 percent of poverty or higher.

Results and Discussion:

**Table 1.** Multiple Comparisons of Student Interest

| Dependent Variable Poverty 1,2,3,4 | Mean Difference | Standard Error | Significance |
|---|---|---|---|
| Student Interest P1–P2, P3–P4 | 11.8000<br>15.6500 | 2.479484<br>2.479484 | .000<br>.000 |
| P2–P1, P3–P4 | -11.8000<br>3.4500 | 2.479484<br>2.479484 | .000<br>.000 |
| P3–P2, P1–P4 | -15.6500<br>-3.4500 | 2.479484<br>2.479484 | .000<br>.000 |
| P4–P2, P3–P1 | 15.6500<br>3.4500 | 2.479484<br>2.479484 | .000<br>.000 |

According to the data summarized in Table 1, the variance for the four poverty levels is statistically significant with respect to student interest. The means were statistically significant in the four different categories at an alpha level of $\alpha= 0.05$. Table 2 displays the data regarding the comparison of extracurricular involvement across the four poverty levels.

**Table 2.** Multiple Comparisons of Extracurriculars

| Dependent Variable Poverty 1,2,3,4 | Mean Difference | Standard Error | Significance |
|---|---|---|---|
| Extracurriculars | | | |
| P1 vs P2 | 10.9400 | 2.479183 | .000 |
| P3 vs P4 | 13.3500 | 2.479183 | .000 |
| P2 vs P1 | -10.9400 | 2.479183 | .000 |
| P3 vs P4 | 1.8100 | 2.479183 | .000 |
| P3 vs P2 | -13.3500 | 2.479183 | .000 |
| P1 vs P4 | -1.8100 | 2.479183 | .000 |
| P4 vs P2 | 13.3500 | 2.479183 | .000 |
| P3 vs P1 | 1.8100 | 2.479183 | .000 |

According to the data summarized in Table 2, the variance for the four poverty levels is statistically significant with respect to student interest. The means were statistically significant in the four different categories at an alpha level of $\alpha = 0.05$. Table 3 displays the data regarding the comparison of student engagement involvement across the four poverty levels.

9**Table 3.** Multiple Comparisons of Student Engagement

| Dependent Variable Poverty 1,2,3,4 | Mean Difference | Standard Error | Significance |
|---|---|---|---|
| Student Engagement | | | |
| P1 — P2 | 12.5500 | 2.479649 | .000 |
| P3 — P4 | 14.1500 | 2.479649 | .000 |
| P2 — P1 | -12.5500 | 2.479649 | .000 |
| P3 — P4 | 2.8200 | 2.479649 | .000 |
| P3 — P2 | -14.1500 | 2.479649 | .000 |
| P1 — P4 | -2.8200 | 2.479649 | .000 |
| P4 — P2 | 14.1500 | 2.479649 | .000 |
| P3 — P1 | 2.8200 | 2.479649 | .000 |

According to the data summarized in Table 3, the variance for the four poverty levels is statistically significant with respect to student interest. The means were statistically significant in the four different categories at an alpha level of $\alpha = 0.05$. Table 4 displays the data regarding the comparison of extracurricular involvement across the four poverty levels.

**Table 4.** Multiple Comparisons of Student Persistence

| Dependent Variable Poverty 1,2,3,4 | Mean Difference | Standard Error | Significance |
|---|---|---|---|
| Student Persistence | | | |
| P1 — P2 | 10.2900 | 2.479679 | .000 |
| P3 — P4 | 13.5300 | 2.479679 | .000 |
| P2 — P1 | -10.2900 | 2.479679 | .000 |
| P3 — P4 | 3.8700 | 2.479679 | .000 |
| P3 — P2 | -12.5300 | 2.479679 | .000 |
| P1 — P4 | -3.8700 | 2.479679 | .000 |
| P4 — P2 | 12.5300 | 2.479679 | .000 |
| P3 — P1 | 3.8700 | 2.479679 | .000 |

Based on the observed means.
The error term is mean square error 45.863.
Tukey HSD

Discussion

The assumptions that were outlined in this study were calculated and the study passed these assumptions tests. The data was collected from the United States Census Bureau and the methods for collecting the data revealed a variety of limitations in the data. The first collection error of the data is that there were strata that were grouped together and the time it took to survey these strata may have affected the data which may not be accounted for in the mean square error. Moreover, the statistics that were presented by the United States Census Bureau was inconsistent because some of the variables were acquired by tabulations of various means and other data points were collection from samples. Moreover, the purpose of collecting and therefore the method used to collect the data points varied from an administrative intention to a purely statistical intention. This may affect the data because there is a higher probability of error within the sample because data that was collected from a survey for statistical purposes may carry a higher percentage of error which may have provoked a vast difference in the variance of the sample gathered for statistical purposes.

Errors that may have affected this study include random and nonrandom nonsampling errors because of the variation in the interpretation of data that exists. Randomness is introduced when people that are surveyed in a study are asked to estimate values which introduce the level of uncertainty and contributes to the inaccuracy of the study. Moreover, the accuracy of the data may be negatively influenced because of bias, inability, difficulty interpreting questions, mistakes in recording data, overcoverage, and undercoverage of data may affect the results that are obtained. These errors contribute largely to the nonrandom uncertainty of the non sampling errors that limit the study. Random nonresponse errors contribute to an overall understatement of the precision of the data. Consequently, new studies would need to be conducted in order to accurately gauge the magnitude of the sampling error empirical estimates.

To correct for the non-sample errors that occurred in this study, the sample estimates are compensated for according to the estimated variance. These adjustments that are made to the model for non sampling errors are referred to as imputations. However, these imputations are limited because it requires a further statistical investigation another statistical investigation to increase the accuracy of the nonsampling error estimate. The typical practice to correct for inaccuracies of imputations is to pair these samples to other samples that have less of a probability to have non sampling errors together that are similar in terms of value. Altogether, this investigation used a standard sample error calculation as estimated by the samples in the study.

Conclusion

The multivariate analysis of variance that was conducted between the income level as defined by poverty status and the student interest in school, amount of extracurricular activities that the student was involved in, the student engagement in the classroom, and the individual student persistence resulted in a conclusive result at 95% confidence or at an alpha level of $\alpha = 0.05$ that there was a statistically significant variance between the poverty level and the activities that students completed as defined by the four categories. This demonstrates that there is a substantial relationship that exists with poverty level and the student academic prosperity. This statistic can be further investigated to find a correlation between a student's level of poverty and involvement in STEM subjects as a means to narrow the research.